\documentclass[12pt]{article}

\usepackage[english]{babel}

\usepackage[letterpaper,top=2cm,bottom=2cm,left=3cm,right=3cm,marginparwidth=1.75cm]{geometry}

\usepackage{amsmath}
\usepackage{graphicx}
\usepackage[colorlinks=true, allcolors=blue]{hyperref} 
\usepackage[utf8]{inputenc} 
\usepackage[T1]{fontenc}    
\usepackage{amsmath}        
\usepackage{amssymb}       
\usepackage{amsfonts}       
\usepackage{graphicx}       
\usepackage{booktabs}       
\usepackage{multirow}       
\usepackage{enumitem}       
\usepackage{url}            
\usepackage{ragged2e}       
\usepackage{tabularx}       
\usepackage{float}      
\usepackage{threeparttable} 
\usepackage{longtable}      
\usepackage{caption}        
\usepackage{subcaption}     
\usepackage{rotating}       
\usepackage{appendix}       
\usepackage{siunitx}        
\usepackage{listings}
\usepackage{xcolor}
\usepackage{rotating}
\usepackage[round,authoryear]{natbib}
\usepackage{setspace}  
\usepackage{authblk}  
\usepackage{pdflscape}
\newcommand{\figurenote}[1]{\par\footnotesize Note: #1} 

\title{ESG Rating Disagreement and Corporate Total Factor Productivity: Inference and Prediction}
\author[1]{Zhanli Li \thanks{Li: Wenlan School of Business, Zhongnan University of Economics and Law, Wuhan, China, 430073; E-mail: lizhanli@stu.zuel.edu.cn.}}
\author[1]{Zichao Yang \thanks{Yang: Wenlan School of Business, Zhongnan University of Economics and Law, Wuhan, China, 430073; E-mail: yang\_zichao@outlook.com, yang\_zichao@zuel.edu.cn}}
\affil[1]{Wenlan School of Business, Zhongnan University of Economics and Law}
\date{}

\begin{document}
\maketitle

\begin{abstract}
This paper examines how ESG rating disagreement ($Dis$) affects corporate total factor productivity (TFP) in China based on data of A-share listed companies from 2015 to 2022. We find that $Dis$ reduces TFP, especially in state-owned, non-capital-intensive, low-pollution and high-tech firms, green innovation strengthens the dampening effect of $Dis$ on TFP, and that $Dis$ lowers corporate TFP by increasing financing constraints and weakening human capital. Furthermore, XGBoost regression demonstrates that $Dis$ plays a significant role in predicting TFP, with SHAP showing that the dampening effect of ESG rating disagreement on TFP is still pronounced in firms with large $Dis$ values.
\end{abstract}

    

\clearpage

\section{Introduction}

Since 1987, the United Nations has championed sustainable development and introduced ESG (Environmental, Social, and Governance) in 2004. Over time, these principles have become key components of global investment strategies and corporate assessments. In recent years, China has embraced sustainable development through initiatives like the philosophy “green mountains and clear waters are as valuable as mountains of gold and silver.” On December 27, 2023, the Central Committee of the Communist Party of China and the State Council issued the \textit{Guidelines for the Comprehensive Advancement of Beautiful China Construction}, which emphasized creating an ESG Evaluation System with Chinese Characteristics.

Despite recent ESG policy efforts, significant disagreements persist within ESG rating systems. Since ESG ratings were introduced in China, multiple agencies have issued their own reports, often resulting in substantial discrepancies. Such differences have considerable implications for evaluated companies. Recent studies highlight the economic impacts of ESG rating disagreement: stock returns suffer \citep{wang2024esg,SDJR202403004}; analyst forecast errors rise \citep{liu2024esg}; green innovation encouragement \citep{HOU2024105914,geng2024esg,ChenPengcheng}; real earnings management increases in both the short and long term \citep{li2024impact}; firms with consistently high ratings benefit from lower costs and increased debt financing \citep{guo2024esg}; disagreement poses a risk of stock price collapse \citep{luo2023duality}; audit fees increase due to information asymmetry \citep{LING2024105749}; and market pressures on managers intensify \citep{CHEN2025104039}.

However, most recent studies have not sufficiently examined ESG rating disagreement’s impact on a company’s overall development. To address this gap, we focus on TFP as an indicator of high-quality development in this study.

\section{Theoretical Analysis and Research Hypotheses}

ESG rating disagreement can confuse stakeholders and increase information asymmetry between companies and investors. While ESG ratings help reduce information asymmetry by providing non-financial insights\citep{ZYCY202307005}, disagreement can amplify asymmetry, creating noise that misleads investors\citep{berg2022aggregate}, raises financing constraints, and ultimately lowers TFP. Stakeholder theory suggests that companies may respond to disagreement by pursuing green innovation\citep{HOU2024105914,geng2024esg,ChenPengcheng}, but this can appear as “greenwashing” to some stakeholders, undermining financing and reducing TFP\citep{PENG2024106012}. Disagreement may also prompt firms to prioritize ESG recognition, causing tension between external investors and internal employees, potentially leading to a human capital outflow and further TFP decline.

Hence, we propose the research hypothesis \textbf{H1:} ESG rating disagreement reduces corporate TFP.
\section{Data and Empirical Design}

\subsection{Sample Selection}
We use financial data of A-share listed companies in China from 2015 to 2022\footnote{Prior 2015, the widely recognized local ESG rating agency in China was Huazheng.}, along with ESG ratings from Huazheng, Wind, SynTao Green Finance, and Susallwave. The data processing steps are as follows: (1) exclude companies in the financial and real estate sectors; (2) exclude ST, *ST, and delisted firms; (3) exclude companies with only one ESG rating in a year; (4) exclude firms with missing key indicators; (5) apply Winsorization at the 1\% level on both tails. The data comes from the CSMAR and CNRDS databases and the four rating agencies, yielding 3656 firms and 13,417 samples in the baseline regression.

\subsection{Model Specification}
To test the hypotheses presented earlier,we conduct the following baseline regression model\footnote{Considering the time required for information to be transmitted and for the market to react and mitigate potential reverse causation problems, TFP is lagged by one period.}:

\begin{equation}
  \label{eq:basic_model}
  TFP_{i,t+1} = \alpha_0 + \alpha_1 Dis_{i,t} + \alpha_2 Control_{i,t} + Year_t + Id_i + \epsilon_{i,t}
\end{equation}
where $TFP_{i,t+1}$ is the total factor productivity of company $i$ in year $t+1$, $Dis_{i,t}$ is the ESG rating discrepancy, $Control_{i,t}$ represents the control variables, $Year_t$ and $Id_i$ denote year and company fixed effects, respectively, and $\epsilon_{i,t}$ is the error term.
\subsection{Variable Descriptions}

(1) Dependent Variable

The dependent variable in this study is Corporate Total Factor Productivity. Considering the robustness of the empirical process, we use three methods to measure TFP, they are LP introduced by\citet{levinsohn2003estimating} and adopted by \citet{lu2012tfp}, OP by \citet{olley1992dynamics} and GMM by \citet{blundell1998initial}. In the prediction task, we use TFP measured by LP method as the dependent variable\citep{XUE}.

\noindent (2) Independent Variable

The independent variable in this study is ESG rating disagreement. Domestic ESG rating agencies have advantages in terms of time and spatial proximity that foreign agencies cannot match, making them more likely to influence the behavior of domestic firms. Therefore, this study uses ESG rating data from four Chinese agencies: Huazheng, Wind, SynTao Green Finance, and Susallwave. Following \citet{Avramov}, the ratings were assigned positive values according to their rating levels, with a difference of 1 between levels. We define the rankings as the values normalized to the range [0, 1] for the given rating agency in a given year, and the pairwise standard deviation of these rankings is used as the measure of $Dis$.
\begin{equation}
Dis_{i,t} = \sqrt{\frac{1}{n-1} \sum_{k \neq j}^{n} (rank_{i,t,k} - rank_{i,t,j})^2} 
\end{equation}
where $rank_{i,j,k}$ represents the normalized ranking of corporate $i$ in year $t$ given by agency $k$, $n$ represents the number of pairs, $k$ and $j$ represent different ESG rating agencies.

\noindent (3) Control Variables

Based on the characteristics of our research and with reference to the works of \citet{YANG2024105850} and \citet{XUE}, the control variables include firm value, top 10 shareholders' ownership percentage, years listed, leverage ratio, firm size, revenue per capita, and ownership ratio. The main variables are defined in \autoref{tab:variable_description}.

\begin{center}
        \textbf{[Insert \autoref{tab:variable_description} Here]}
\end{center}
\section{Empirical Analysis}

\subsection{Baseline Regression}
\autoref{tab:baseline_regression} reports the results of the baseline regression, with clustering at the individual and time levels. The results indicate that ESG rating disagreement reduces corporate TFP, providing preliminary support for the validity of Hypothesis H1.

\begin{center}
        \textbf{[Insert \autoref{tab:baseline_regression} Here]}
\end{center}

\subsection{Mechanism Analysis}

\subsubsection{The Role of Green Innovation}
Myriad existing researches demonstrate that ESG rating disagreement can drive firms to enhance green innovation\citep{HOU2024105914,geng2024esg,chen2024esg}. However, does this green innovation truly contribute to substantial corporate development? This study introduces green innovation as a mechanism variable, measured by the natural logarithm of the number of green patents held by a company. To examine the role of green innovation in the relationship between ESG rating disagreement and TFP, an interaction term between green innovation and ESG rating disagreement is constructed, with other settings remaining consistent with \autoref{eq:basic_model}.

\begin{figure}[H]
            	\centering
            	\begin{minipage}{0.32\textwidth}
            		\centering
            		\includegraphics[width=\textwidth]{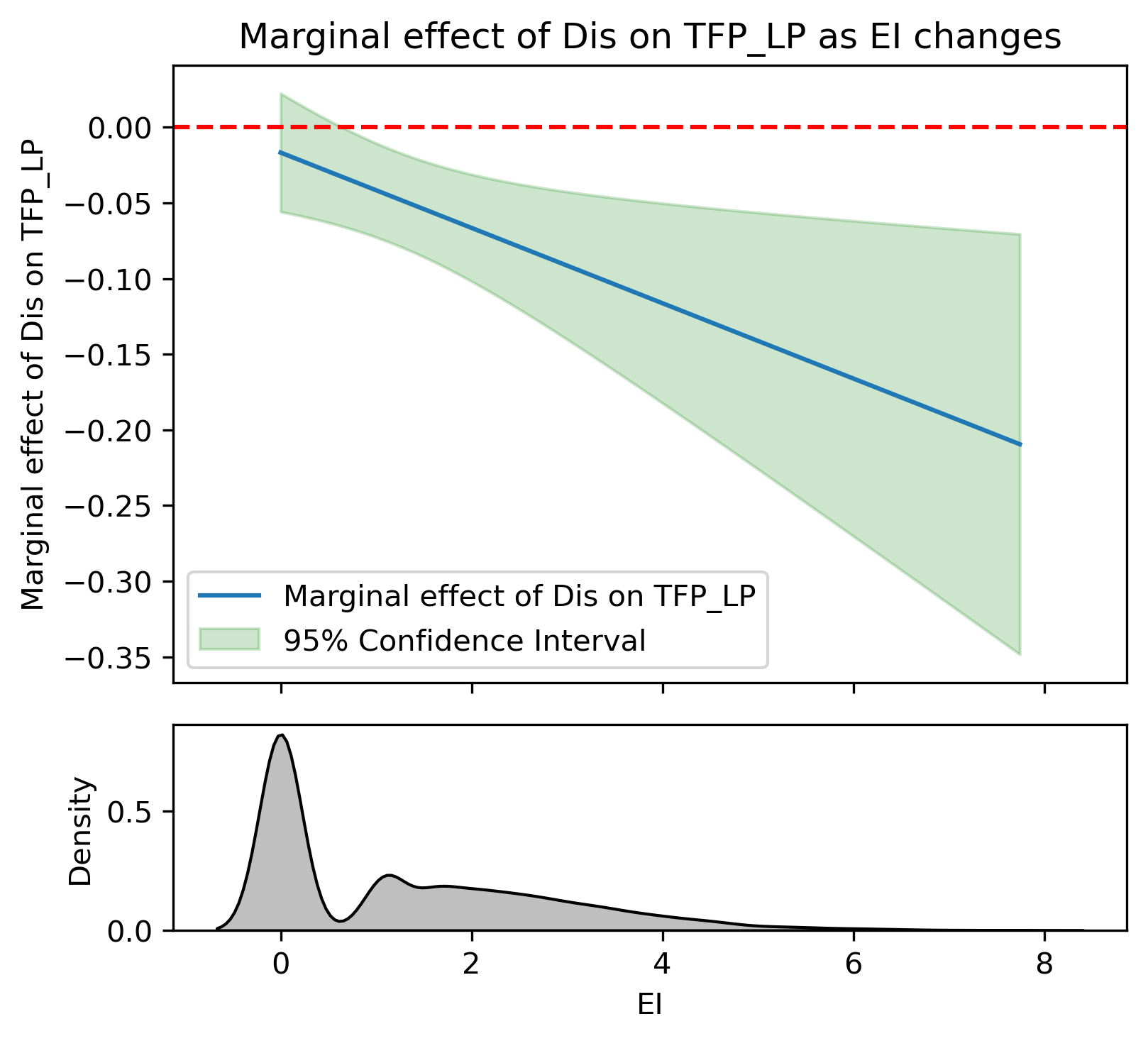}
            	\end{minipage}
            	\hfill
            	\begin{minipage}{0.32\textwidth}
            		\centering
            		\includegraphics[width=\textwidth]{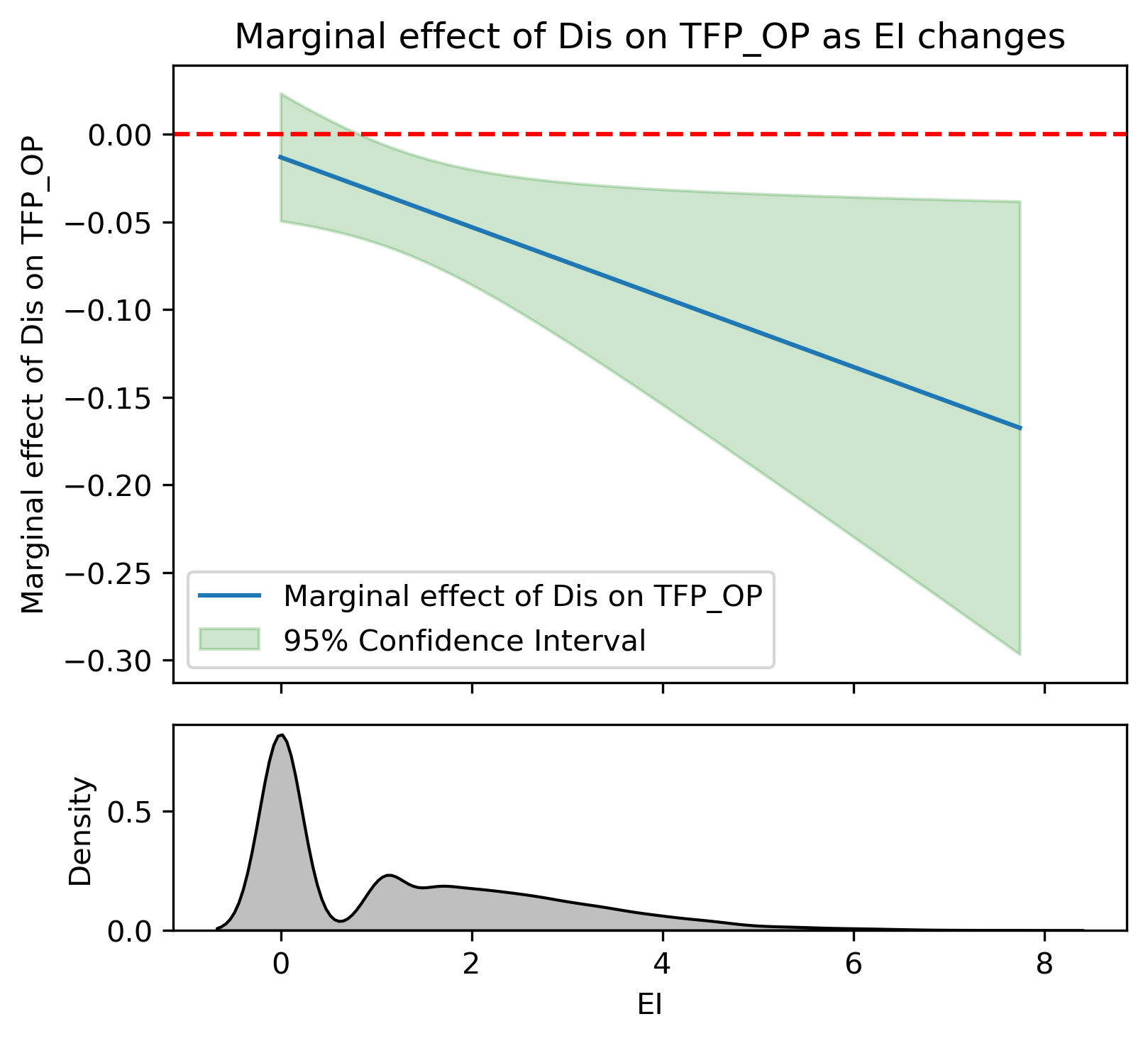}
            	\end{minipage}
            	\hfill
            	\begin{minipage}{0.32\textwidth}
            		\centering
            		\includegraphics[width=\textwidth]{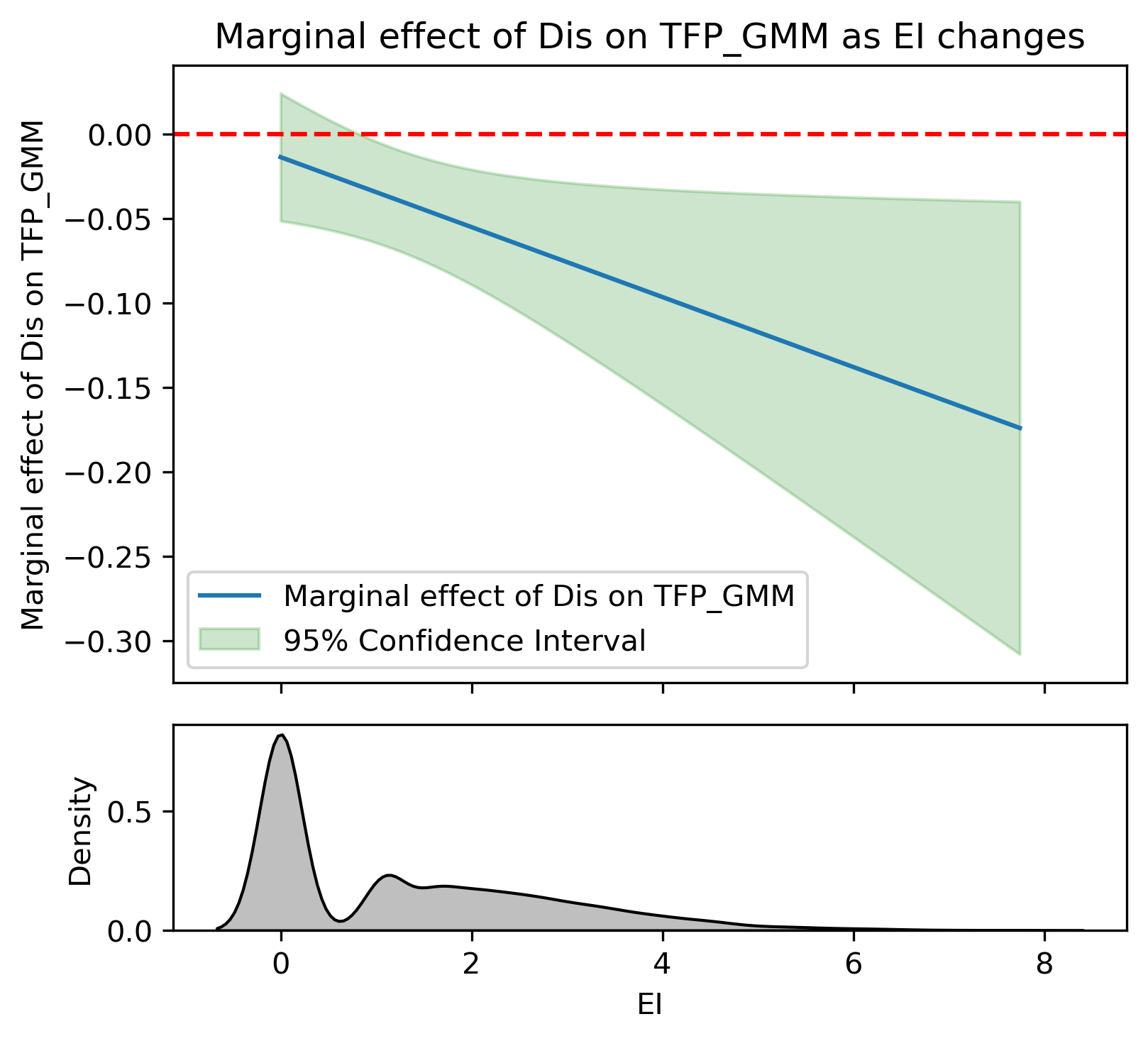}
            	\end{minipage}
            	\caption{Marginal Effect}
                \figurenote{\raggedright This figure visualize the impact of green innovation on the marginal effect of ESG rating disagreement on TFP, verifying that green innovation exacerbates this negative effect.}
            	\label{fig:mc_plt}
            \end{figure}

\autoref{tab:interaction} and \autoref{fig:mc_plt} report the results. The results show that when there are no ESG rating discrepancies, green innovation promotes the improvement of TFP. Conversely, when there are ESG rating discrepancies, it strengthens the dampening effect of ESG rating disagreement on TFP. The result seem contradict exiting researches, which argue ESG rating disagreement can drive firms to enhance green innovation\citep{HOU2024105914,geng2024esg,chen2024esg}. This observation is very likely caused by a remedial action taken by firms in response to ESG rating disagreement, as \citet{HOU2024105914} mentioned, such ESG rating disagreement is negatively correlated with the quality of corporate green innovation. According to stakeholder theory, firm's remedial green innovations may be seen as "greenwashing" when there is disagreement over ESG ratings, which will lead to higher costs of debt financing for corporations, ultimately bringing about a decline in TFP\citep{PENG2024106012}. 

\begin{center}
        \textbf{[Insert \autoref{tab:interaction} Here]}
\end{center}

\begin{center}
        \textbf{[Insert \autoref{fig:mc_plt} Here]}
\end{center}

To further investigate whether the moderating effect varies across levels of ESG rating disagreement, we conducted regressions separately for the upper and lower 50\% quartiles of ESG rating disagreement, and the \autoref{tab:interaction2} show that the moderating effect is significant only when the ESG rating disagreement is high, further supporting the notion that substantial ESG rating disagreement may lead to stakeholder distrust in firms’ green innovation.
\begin{center}
        \textbf{[Insert \autoref{tab:interaction2} Here]}
\end{center}

\subsubsection{The Role of Financing Constraint}
Undoubtedly, ESG investment plays a crucial role in promoting ESG development, making it essential to examine the impact of ESG rating disagreement on corporate financing. According to the information asymmetry theory discussed in the previous section, ESG rating disagreement introduces noise, potentially contributes financing constraints for firms. This study focuses exclusively on data from domestic ESG rating agencies in China, in contrast to \citet{Li}, which also includes data from foreign rating agencies to explore this mechanism. Hence, we employ the WW and KZ indices form CSAMR to enhance the robustness of the empirical evidence. Furthermore, we follow \citet{JiangTing}'s research on the transmission mechanism, to avoid financing constraints being both exogenous and endogenous. We only use independent variable to regress with finacing constraints, since numerous researches have demonstrated that financing constraints are detrimental to TFP improvement\citep{hopenhayn2014firms,zhang2021financial,piao2023financial}. 

\autoref{tab:procuration} reports the regression results. The $Disb$ is the variable by one period. So we concludes that increased ESG rating disagreement strengthens corporate financing constraints, thereby lowering TFP.
\begin{center}
        \textbf{[Insert \autoref{tab:procuration} Here]}
\end{center}
\subsubsection{The Role of Human Capital}
Human capital is widely acknowledged as a positive contributor to TFP\citep{MILLER2000399,xu2008openness,HUANG2024109901}. Employees with a bachelor's or master's degree typically reflects higher levels of knowledge, skills, and expertise among workers, enabling them to perform task more effciently, reduce errors, and enhance productivity. In this subsection, we delve deeper into how ESG rating disagreement interacts with human capital and its potential impact on a company’s TFP, with a specific focus on employees holding bachelor’s or master’s degrees. Building on \citet{JiangTing}'s research on the transmission mechanism, and following a similar approach to verifying the financing constraint mechanism to avoid endogeneity issues in mediating variables, we perform regressions of $Dis$ on $Master$ and $Dis$ on $Undergrad$.

\autoref{tab:human} reports the regression results, showing that \textit{Dis} is detrimental to the development of firms' human capital, 
which serves as a crucial driver of TFP. Based on this evidence, we argue that \textit{Dis} reduces TFP by impairing human capital.
\begin{center}
        \textbf{[Insert \autoref{tab:human} Here]}
\end{center}

\subsection{Heterogeneity Analysis}
To further explore whether the dampening effect of ESG rating disagreement on TFP exhibits heterogeneity, this study conducts a subgroup regression analysis. Given that the sample size decreases after subgrouping, making the results more susceptible to the influence of outliers, the data is Winsorized at the 2\% level on both ends before performing the subgroup regression and conducting a Chow Test on $Dis$. 

\autoref{tab:Hethero} reports the results of the heterogeneity analysis. The results indicate that the effect is more significant in state-owned firms, non-capital-intensive firms, low-pollution firms and high-tech firms. This heterogeneity may arise for the following reasons: first, non-state-owned firms are generally more willing to engage in ESG initiatives as they tend to focus more on ESG performance to attract investment; second, capital-intensive firms are usually less influenced by market forces due to their low variable costs and high fixed costs, which buffer them from the pressures of ESG-related changes; third, highly polluting firms, which are often subject to stringent environmental policies and regulations in their financing, are less likely to invest in ESG performance, making the impact of divergent ESG ratings less pronounced. And, finally, as previously analyzed, ESG rating disagreements can lead to the loss of highly qualified employees, on whom high-tech firms are particularly dependent. Consequently, the negative impact of disagreements on TFP is more significant for high-tech firms.
\begin{center}
        \textbf{[Insert \autoref{tab:Hethero} Here]}
\end{center}

\section{Further Discussion}
In the above analysis we can see that ESG rating disagreement brings about a decrease in TFP, which is obtained based on linear regression. In order to further explore the relationship between ESG rating divergence and TFP under the nonlinear assumption, we consider a regression analysis using a machine learning approach that can be interpreted with SHapley Additive exPlanations(SHAP) introduced by \citet{scott2017unified}, which can observe whether ESG rating disagreement is crucial for explaining TFP under the assumption of nonlinear system and explore the nonlinear influence relationship between the two variables.

Following \citet{XUE}, we use the XGBoost machine learning model introduced by \citet{chen2016XGBoost} to predict $TFP_{LP}$. XGBoost is an ensemble learner based on gradient boosted trees (GBDT) that has demonstrated excellent performance in many applications, even outperforming deep neural networks in tabular data\citep{grinsztajn2022tree}. In this paper, the Optuna library developed by \citet{akiba2019optuna}, combed with early stopping and cross-validation strategies is employed for parameter selection. After that, we adopt the SHAP (SHapley Additive exPlanations) method proposed by \citet{scott2017unified} to explain XGBoost. The variables include the ESG rating disagreement ($Dis$ \& $Disb$), all control variables, and institutional ownership percentage ($Inst$).

We use 80\% of the dataset for training and 20\% for testing, conducting three rounds of training. The first round includes ESG rating disagreement (\textit{Dis}), the second round incorporates ESG rating disagreement from the previous period, 
and the third round excludes ESG rating disagreement altogether. \autoref{tab:calsusal} presents the evaluation metrics for these three rounds of training.

First, compared to the linear model (Set4), XGBoost demonstrates significantly better performance across all metrics, 
highlighting the machine learning model's ability to capture nonlinear relationships. 
Second, when comparing XGBoost without incorporating \textit{Dis} (Set3), 
the model's performance improves by 4.3\% (Set1) and 6.23\% (Set2), respectively, using $R^2$ as the evaluation criterion. 
This improvement underscores the predictive value of ESG ratings in modeling TFP.
\begin{center}
        \textbf{[Insert \autoref{tab:calsusal} Here]}
\end{center}

To analyze the influence of ESG rating disagreement on $TFP_{LP}$ in our predictive task, we use SHAP to interpret the results of the XGBoost model. SHAP values are grounded in cooperative game theory and provide a way to understand each feature's contribution to the model output by treating features as "players" in a game. The contribution, or Shapley value $\phi_i$, for a particular feature $i$ is computed by averaging its marginal contributions across all possible combinations of features.

Mathematically, the Shapley value $\phi_i$ for feature $i$ is defined as:
\begin{equation}
\phi_i = \sum_{S \subseteq N \setminus \{i\}} \frac{|S|! \cdot (|N| - |S| - 1)!}{|N|!} \left[ f(S \cup \{i\}) - f(S) \right]
\end{equation}
where $N$ represents the set of all features, $S \subseteq N \setminus \{i\}$ is a subset excluding feature $i$, $|S|$ denotes the cardinality of subset $S$, $f(S)$ indicates the model output using feature subset $S$, and $f(S \cup \{i\})$ represents the output when feature $i$ is added to $S$.

The term $\frac{|S|! \cdot (|N| - |S| - 1)!}{|N|!}$ in the equation represents a weighting factor that accounts for all possible orders in which features can be added to the model. By calculating $\phi_i$ across all possible subsets $S$, SHAP captures the average marginal effect of adding feature $i$ to different subsets of features, thereby quantifying the feature's contribution to the model’s output.

\begin{figure}[H]
                \centering
                \includegraphics[width=0.85\textwidth]{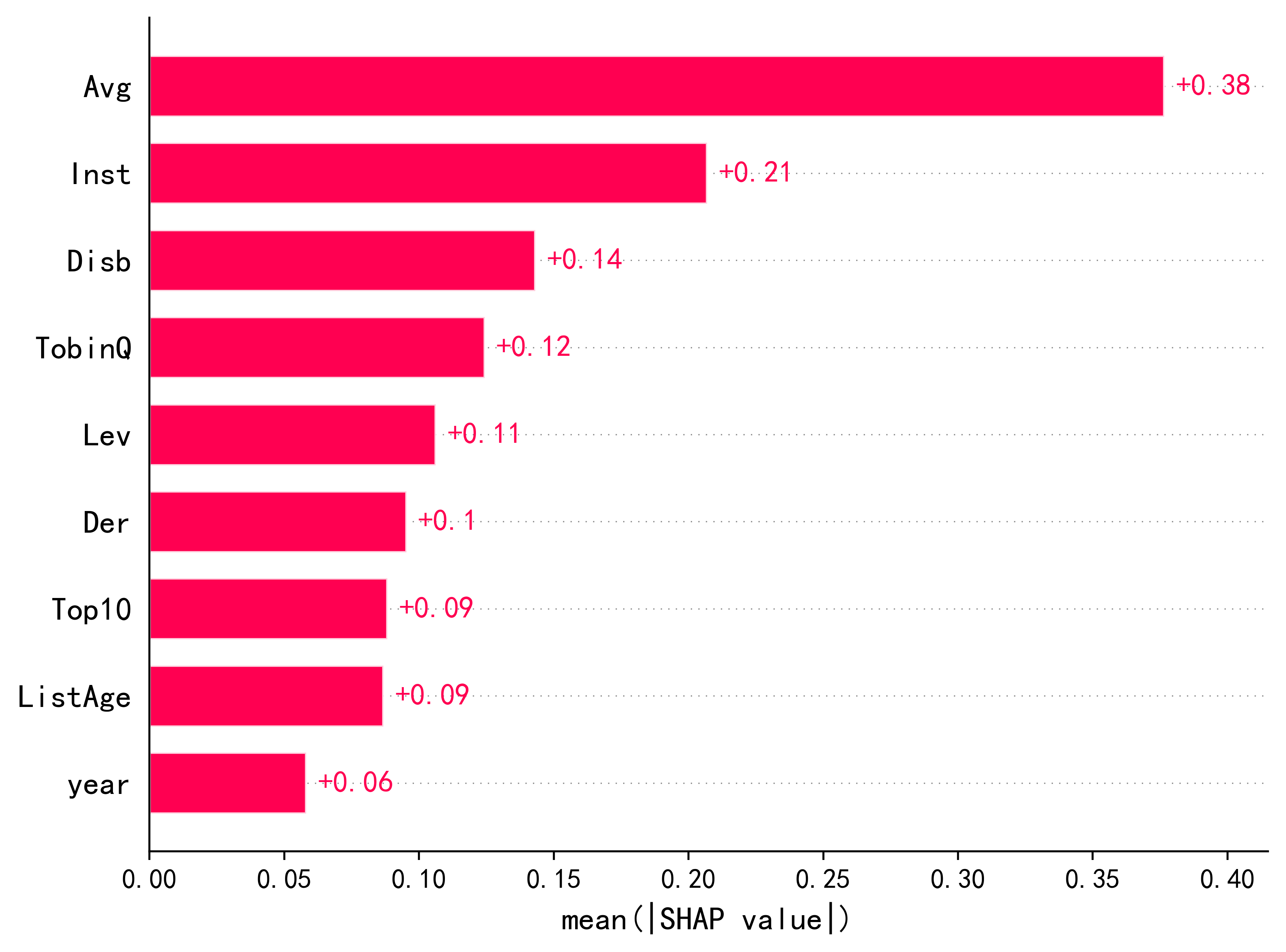}
                \caption{Feature Importance}
                \figurenote{\raggedright The importance of each feature is determined by the percentage ratio of its average absolute SHAP value to the mean absolute SHAP value of all features. A higher percentage indicates that the feature is more influential relative to the others.}
                \label{fig:feature_importance}
            \end{figure}

\begin{figure}[H]
                \centering
                \includegraphics[width=0.85\textwidth]{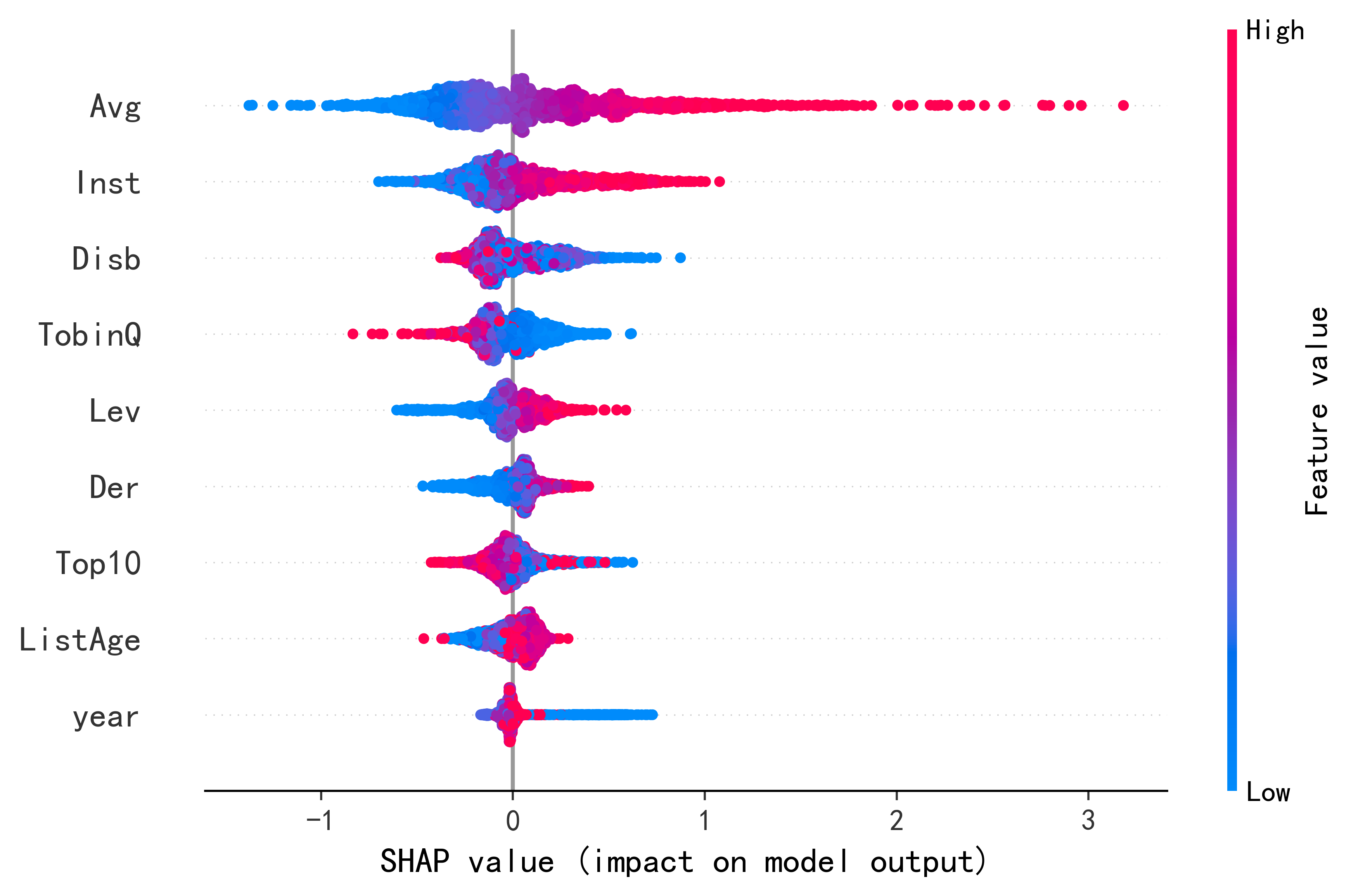}
                \caption{SHAP Beeswarm Plot}
                \figurenote{\raggedright The beeswarm plot illustrates how different feature values impact TFP. The color of the dots transitions from red (for larger eigenvalues) to blue (for smaller eigenvalues). Dots positioned to the left of 0 on the horizontal axis indicate a negative effect on the dependent variable, while those to the right of 0 represent a positive effect.}
                \label{fig:shap_beeswarm}
            \end{figure}

In this study, based on the previous training results, we chose the model with one-period lagged ESG rating disagreement for our SHAP analysis, and the SHAP values were calculated for the test set. \autoref{fig:feature_importance} reports the feature importance, showing that the contribution ratio of $Disb$ is 0.14, again indicating that $Disb$ is crucial for predicting TFP. \autoref{fig:shap_beeswarm} reports the SHAP beeswarm plot, it can be observed that high ESG rating divergence suppresses a company's TFP, while in companies with low ESG ratings and small divergences, ESG rating disagreement is even positively correlated with TFP in the prediction task. The reason for this phenomenon is that the investment market is more tolerant of small ESG rating divergence, and listed companies will not face significant financing pressure due to small divergence. 
\begin{center}
        \textbf{[Insert \autoref{fig:feature_importance} Here]}
\end{center}

\begin{center}
        \textbf{[Insert \autoref{fig:shap_beeswarm} Here]}
\end{center}

\section{Conclusion and Implication}
ESG rating disagreement presents significant challenges to the improvement of corporate TFP, underscoring the urgent need for firms to address this issue. Mechanism analysis reveals that green innovation can exacerbate the negative effects of ESG rating disagreement on TFP. Additionally, these disagreement hinder TFP growth by intensifying financing constraints and weakening human capital. To mitigate these impacts, companies should focus on improving the quality of ESG information disclosure,  carefully navigate the trade-offs between green innovation, financing needs, and strive to balance the interests of diverse stakeholders. Regulatory authorities, in turn, should strengthen the scrutiny of green innovation patents to ensure their quality and prevent superficial compliance efforts.

Heterogeneity analysis demonstrates that the negative effects of ESG rating disagreement on TFP are more pronounced in state-owned enterprises, non-capital-intensive firms, low-pollution industries, and high-tech enterprises. These entities should pay closer attention to ESG rating discrepancies, while regulators should provide targeted support to low-pollution and high-tech firms to foster sustainable development and technological innovation. Moreover, ESG rating disagreement serve as a critical predictor of TFP, with SHAP values highlighting their substantial and ongoing negative impact. This indicator can be leveraged by listed companies and regulatory agencies as a monitoring and early warning tool to identify potential risks to TFP and implement proactive measures.

We believe it would be worthwhile to further investigate how green innovation amplifies the negative impact of ESG rating divergence on TFP, as well as to integrate SHAP with microeconometric methods for more robust empirical analysis.
\newpage
\bibliography{frl}
\newpage

\section{Tables and Figures in the Main Text}
\begin{table}[H]
        \centering
        \caption{Definition of Main Variables}
        \label{tab:variable_description}
        \begin{threeparttable}

        \begin{tabular}{c m{4cm} p{9.4cm}}
        \toprule
        \textbf{Symbol} & \textbf{Source} & \textbf{Description} \\
        \midrule
        \multirow{1}{*}{$TFP$} & \multirow{1}{*}{CSAMR} & Estimated using the LP, OP and GMM method \\
        \multirow{1}{*}{$Dis$} &   \multirow{1}{*}{ESG Rating Agencies} & Calculated based on the method described above \\
        $TobinQ$ & CSAMR & Tobin's Q ratio\\
         \multirow{1}{*}{$Top10$} &   \multirow{1}{*}{CSAMR} & Shareholding of top 10 shareholders / total shares\\
         \multirow{1}{*}{$ListAge$} &   \multirow{1}{*}{CSAMR} & Natural logarithm of years since the firm's IPO\\
         \multirow{1}{*}{$Lev$} &   \multirow{1}{*}{CSAMR} & Total liabilities at year-end / total assets at year-end\\
         \multirow{1}{*}{$Size$} &   \multirow{1}{*}{CSAMR} & Natural logarithm of total assets at year-end\\
         \multirow{2}{*}{$Avg$} &   \multirow{2}{*}{CSAMR} & Natural logarithm of operating income / number of employees\\
          \multirow{1}{*}{$Der$} &   \multirow{1}{*}{CSAMR} & Total liabilities at year-end / total equity at year-end\\
        \multirow{2}{*}{$EI$} &   \multirow{2}{*}{CNRDS} & Natural logarithm of the number of green patents obtained in a given year\\
        \multirow{2}{*}{$FC$} &   \multirow{2}{*}{CSAMR} & WW and KZ indices, the larger they are the more constrained the financing is\\
        \multirow{3}{*}{$HC$} &   \multirow{3}{*}{CSAMR} & Natural logarithm of number of graduate students and above and natural logarithm of number of undergraduate students and above\\
        \bottomrule
        \end{tabular}
        \begin{tablenotes}
                \item Note: This table presents the main variables used in the paper.
        \end{tablenotes}
\end{threeparttable}
      \end{table}
\begin{table}[H]
        \centering
        \caption{Baseline Regression Results}
        \label{tab:baseline_regression}
        \begin{threeparttable}
        \begin{tabular}{lcccccc}
        \toprule
        \textbf{Variables} & \multicolumn{6}{c}{\textbf{TFP}}\\
        \cmidrule(lr){2-7}
        & \textbf{$TFP_{LP}$} & \textbf{$TFP_{OP}$} & \textbf{$TFP_{GMM}$} & \textbf{$TFP_{LP}$}& \textbf{$TFP_{OP}$}& \textbf{$TFP_{GMM}$}\\
        \midrule
        \textbf{$Dis$}     & -0.0675*** & -0.0536** & -0.0536** & -0.0447*** & -0.0357** & -0.0373** \\
                & (-2.7452)  & (-2.1838) & (-2.2098) & (-2.9449)  &  (-2.2503) & (-2.2395) \\
        \textbf{$TobinQ$}  &           &           &           & -0.0032     & -0.0017  & -0.0007  \\
                &           &           &           & (-0.5471)   & (-0.3953) & (-0.1795) \\
        \textbf{$Top10$}   &           &           &           & -0.0010    & -0.0009 & -0.0005 \\
                &           &           &           & (-0.9976)  & (-0.9496) & (-0.4838) \\
        \textbf{$ListAge$}  &           &           &           & 0.0989**  & 0.0855** & 0.0160 \\
                &           &           &           & (2.4952)   & (2.3667) & (0.4241) \\
        \textbf{$Lev$}    &           &           &           & -0.1697  & -0.1557 & -0.1481 \\
                &           &           &           & (-1.5121)  & (-1.5361) & (-1.4348) \\
        \textbf{$Size$}     &           &           &           & 0.3737***  & 0.2581*** & 0.2245*** \\
                &           &           &           & (13.129)   & (14.214) & (10.939) \\
        \textbf{$Avg$}     &           &           &           & 0.2292***  & 0.2695*** & 0.2671*** \\
                &           &           &           & (2.7852)   & (2.8274) & (2.7426) \\
        \textbf{$Der$}     &           &           &           & -0.0210**  & -0.0190* & -0.0211** \\
                &           &           &           & (-2.0405)  & (-1.8342) & (-1.9816) \\
        \midrule
        \textbf{Controls} & NO & NO & NO & YES & YES & YES \\
        \textbf{Year} & YES     & YES & YES & YES & YES & YES \\
        \textbf{Id} & YES     & YES & YES & YES  & YES & YES \\
        \textbf{Doubel Clustering}& YES     & YES & YES & YES  & YES & YES \\
        \textbf{$N$}   & 13417     & 13417 & 13417 & 13417  & 13417 & 13417 \\
        \textbf{$R^2$}    & 0.0015    & 0.0011 & 0.0011 & 0.1785  & 0.1670 & 0.1409 \\
        \bottomrule
        \end{tabular}
        \begin{tablenotes}
        \item Note: This table presents the benchmark regression results, confirming the negative impact of ESG rating disagreement on TFP. T-statistics are shown in parentheses. * p<0.1, ** p<0.05, *** p<0.01. The same applies below.
        \end{tablenotes}
        \end{threeparttable}
      \end{table}
      \begin{table}[H]
        \centering
        \caption{Green Innovation Mechanism}
        \label{tab:interaction}
        \begin{threeparttable}
        \begin{tabular}{lcccccc}
        \toprule
        \textbf{Variables} & \multicolumn{6}{c}{\textbf{TFP}} \\
        \cmidrule(lr){2-7}
        & \textbf{$TFP_{LP}$} & \textbf{$TFP_{OP}$}&\textbf{$TFP_{GMM}$} & \textbf{$TFP_{LP}$} & \textbf{$TFP_{OP}$} & \textbf{$TFP_{GMM}$} \\
        \midrule
        \textbf{$Dis$} & -0.0264 & -0.0205& -0.0199& -0.0167 & -0.0130 & -0.0136 \\
        & (-1.2045) &(-1.0114) & (-0.9590)& (-0.8367) & (-0.7014) & (-0.7069) \\
        \textbf{$EI$} & 0.0412*** &0.0283*** & 0.0253***& 0.0223*** & 0.0142*** & 0.0126*** \\
        & (8.4064) &(6.2383) &(5.4522) & (4.9596) & (3.3848) & (2.8967) \\
        \textbf{$Dis*EI$} & -0.0366*** &-0.0292*** &-0.0296*** & -0.0249** & -0.0199** & -0.0207** \\
        & (-3.1706) &(-2.7348) & (-2.7184)& (-2.3693) & (-2.0401) & (-2.0432) \\
        \textbf{$TobinQ$} &  & & & -0.0030 & -0.0016 & -0.0007 \\
        &  & & & (-0.7895) & (-0.4611) & (-0.1778) \\
        \textbf{$Top10$} &  & & & -0.0010* & -0.0010* & -0.0005 \\
        &  & & & (-1.7413) & (-1.7832) & (-0.8682) \\
        \textbf{$ListAge$} &  & & & 0.1012*** & 0.0869*** & 0.0172 \\
        &  & & & (3.4856) & (3.2139) & (0.6122) \\
        \textbf{$Lev$} &  & & & -0.1692*** & -0.1550*** & -0.1472*** \\
        &  & & & (-3.2834) & (-3.2318) & (-2.9594) \\
        \textbf{$Size$} &  & & & 0.3664*** & 0.2537*** & 0.2208*** \\
        &  & & & (28.243) & (21.017) & (17.632) \\
        \textbf{$Avg$} &  & & & 0.2297*** & 0.2698*** & 0.2673*** \\
        &  & & & (22.666) & (28.608) & (27.331) \\
        \textbf{$Der$} &  & & & -0.0210*** & -0.0190*** & -0.0211*** \\
        &  & & & (-3.1573) & (-3.0646) & (-3.2805) \\
        \midrule
        \textbf{Controls} & NO & NO& NO& YES & YES & YES \\
        \textbf{Year} & YES & YES&YES & YES & YES & YES \\
        \textbf{Id} & YES &YES &YES & YES & YES & YES \\
        \textbf{$N$} & 13417 & 13417&13417 & 13417 & 13417 & 13417 \\
        \textbf{$R^2$} & 0.0092 &0.0052 & 0.0042& 0.1806 & 0.1680 & 0.1417 \\
        \bottomrule
        \end{tabular}
        \begin{tablenotes}
                \item Note: This table presents the results of the interaction term regression for green innovation, demonstrating that under ESG rating disagreement, green innovation contributes to a decline in TFP.
        \end{tablenotes}
        \end{threeparttable}
    \end{table}
    
    \begin{table}[H]
        \centering
        \caption{Heterogeneity of Green Innovation Mechanism}
        \label{tab:interaction2}
        \begin{threeparttable}
        \begin{tabular}{lcccccc}
        \toprule
    \textbf{Variables} & \multicolumn{3}{c}{\textbf{High $Dis$}}& \multicolumn{3}{c}{\textbf{Low $Dis$}}\\
        \cmidrule(lr){2-4} \cmidrule(lr){5-7}
        & \textbf{$TFP_{LP}$} & \textbf{$TFP_{OP}$}&\textbf{$TFP_{GMM}$} & \textbf{$TFP_{LP}$} & \textbf{$TFP_{OP}$} & \textbf{$TFP_{GMM}$} \\
        \midrule
        \textbf{$Dis$} & 0.0527& 0.0616& 0.0501& -0.2268***& -0.1532**& -0.1554**\\
        & (1.2360)&(1.5631)& (1.2368)& (-2.8516)& (-2.0471)& (-1.9903)\\
        \textbf{$EI$} & 0.0364***&0.0280***& 0.0230**& 0.0092& 0.0043& 0.0041\\
        & (3.1495)&(2.6193)&(2.0914)& (1.4601)& (0.7258)& (0.6628)\\
        \textbf{$Dis*EI$} & -0.0522**&-0.0461**&-0.0381**& 0.0433& 0.0280& 0.0260\\
        & (-2.1556)&(-2.0558)& (-1.6525)& (1.2832)& (0.8798)& (0.7855)\\
        \textbf{$TobinQ$} &  -0.0009& -0.0012& 0.0020& -0.0025& -0.0020& -0.0013\\
        &  (-0.1395)& (0.1681)& (0.3447)& (-0.4475)& (-0.3853)& (-0.2318)\\
        \textbf{$Top10$} &  -0.0008& -0.0012& -0.0005& -0.0001& 0.0004& 0.0008\\
        &  (-0.8389)& (-1.3029)& (-0.5405)& (-0.1141)& (0.4905)& (0.8934)\\
        \textbf{$ListAge$} &  0.1137**& 0.0833*& 0.0257& 0.0609& 0.0529& -0.0332\\
        &  (2.2567)& (1.7886)& (0.5359)& (1.4289)& (1.3186)& (-0.7922)\\
        \textbf{$Lev$} &  -0.1140& -0.0997& -0.0976& -0.2786***& -0.2703***& -0.2712***\\
        &  (0.1716)& (-1.2930)& (-1.2306)& (-3.2945)& (-3.3962)& (-3.2670)\\
        \textbf{$Size$} &  0.3607***& 0.2525***& 0.2173***& 0.3523***& 0.2355***& 0.2123***\\
        &  (16.816)& (12.732)& (10.655)& (17.502)& (12.434)& (10.746)\\
        \textbf{$Avg$} &  0.2126**& 0.2424***& 0.2447***& 0.2301***& 0.2706***& 0.2651***\\
        &  (13.198)& (16.277)& (15.983)& (13.969)& (17.451)& (16.389)\\
        \textbf{$Der$} &  0.0224**& -0.0210**& -0.0226**& 0.0046& 0.0052& 0.0036\\
        &  (-2.2377)& (-2.2735)& (-2.3751)& (0.3638)& (0.4404)& (0.2910)\\
        \midrule
        \textbf{Controls} & YES& YES& YES& YES & YES & YES \\
        \textbf{Year} & YES & YES&YES & YES & YES & YES \\
        \textbf{Id} & YES &YES &YES & YES & YES & YES \\
        \textbf{$N$} & 6774& 6774&6774 & 6643& 6643 & 6643 \\
        \textbf{$R^2$} & 0.1578&0.1434& 0.1229& 0.1717& 0.1533& 0.1310\\
        \bottomrule
        \end{tabular}
        \begin{tablenotes}
                \item Note: This table displays the subgroup regression results, illustrating the heterogeneity in the moderating effects of green innovation.
        \end{tablenotes}
        \end{threeparttable}
    \end{table}
    \begin{landscape}
        \begin{table}[H]
            \centering
            \caption{Financing Constraints Mechanism}
            \small
            \label{tab:procuration}
            \begin{threeparttable}
            \begin{tabular}{lcccccccc}
            \toprule
            \textbf{Variables} &  &\multicolumn{7}{c}{\textbf{Financing Constraints}}\\
            \cmidrule(lr){2-9}& \textbf{$WW$} & \textbf{$WW$}&\textbf{$KZ$}&  \textbf{$KZ$}&\textbf{$WW$}  & \textbf{$WW$}&\textbf{$KZ$} &\textbf{$KZ$}\\
            \midrule
            \textbf{$Dis$}     & &0.0122***& &  0.2166***&& 0.0089***& &0.1815***\\ 
                        & &(4.7408)& &  (3.0974)&& (3.6730)& &(2.6598)\\
            \textbf{$Disb$}     & 0.0122***&& 0.2142***&  &0.0074***& & 0.1392**&\\ 
                    & (4.7691)&& (3.0604)&  &(3.3430)& & (2.2547)&\\
            \textbf{$TobinQ$}  &     &      & &  &-0.0022***& -0.0006& 0.2724***&0.0237\\
                    &   &        & &  &(-3.9269)& (-0.9530)& (19.528)&(1.4433)\\
            \textbf{$Top10$}   &    &       & &  &-0.0007***& -0.0006***& 0.0011&-0.0061**\\
                    &       &    & &  &(-8.8614)& (-6.9840)& (0.4783)&(-2.3496)\\
            \textbf{$ListAge$}  &  &         & &  &0.0204***& 0.0143***& 0.5530***&1.2048***\\
                    &      &     & &  &(6.1906)& (2.9544)& (6.5333)&(9.5882)\\
            \textbf{$Lev$}     &   &        & &  &0.0370***& 0.0359***& 6.7446***&3.1453***\\
                    &      &     & &  &(5.0820)& (4.5416)& (32.842)&(14.109)\\
            \textbf{$Size$}    &     &      & &  &-0.0614***& -0.0455***& -0.4512***&-0.2274***\\
                    &    &       & &  &(-36.832)& (-22.973)& (-9.5683)&(-4.0781)\\
            \textbf{$Avg$}     &     &      & &  &-0.0260***& -0.0113***& -0.5288***&-0.0107\\
                    &      &     & &  &(-18.835)& (-7.3282)& (-13.531)&(-0.2432)\\
            \textbf{$Der$}     &     &      & &  &0.0030***& 0.0011& 0.0191&-0.0027\\
                    &      &     & &  &(3.4646)& (1.2994)& (0.6820)&(-0.0931)\\
            \midrule
            \textbf{Controls} & NO &NO& NO&  NO&YES   & YES& YES  &YES\\
            \textbf{Year} & YES  &YES   & YES       &  YES&YES  & YES& YES  &YES\\
            \textbf{Id} & YES  &YES   & YES       &  YES&YES  & YES& YES  &YES\\
            \textbf{$N$}   & 10059&10059& 13417&  13417&10059& 10059& 13417&13417\\
            \textbf{$R^2$}    & 0.0034&0.0034& 0.0010&  0.0010&0.2647& 0.1221& 0.2250&0.0517\\
            \bottomrule
            \end{tabular}
            \begin{tablenotes}
            \item Note: Similar to the robustness tests, $Dis$ corresponds to regressions on ESG rating disagreement in year t and financing constraints in year t+1, while $Disb$ corresponds to regressions on ESG rating disagreement in year t-1 and financing constraints in year t.
            \end{tablenotes}
            \end{threeparttable}
        \end{table}
        \end{landscape}
    \begin{landscape}
        \begin{table}[H]
        \centering
        \caption{Human Capital Mechanism}
        \small
        \label{tab:human}
        \begin{threeparttable}
        \begin{tabular}{lcccccccc}
        \toprule
        \textbf{Variables} &  &\multicolumn{7}{c}{\textbf{Human Capital}}\\
        \cmidrule(lr){2-9}& \textbf{$Master$} & \textbf{$Master$}&\textbf{$Undergrad$}&  \textbf{$Undergrad$}&\textbf{$Master$}  & \textbf{$Master$}&\textbf{$Undergrad$} &\textbf{$Undergrad$}\\
        \midrule
        \textbf{$Dis$}     & &-0.0595***& &  -0.0526***&& -0.0345**& &-0.0316**\\ 
                    & &(-2.8695)& &  (-3.1495)&& (-1.7625)& &(-2.0461)\\
        \textbf{$Disb$}     & -0.0598***&& -0.0528**&  &-0.0322*& & -0.0248*&\\ 
                & (-2.8867)&& (-3.1608)&  &(-1.8268)& & (-1.9302)&\\
        \textbf{$TobinQ$}  &     &      & &  &0.0197***& -0.0067& 0.2724***&-0.0036\\
                &   &        & &  &(5.0228)& (-1.4155)& (19.528)&(-0.9603)\\
        \textbf{$Top10$}   &    &       & &  &0.0029***& 0.0008& 0.0011&0.0003\\
                &       &    & &  &(4.4620)& (1.0079)& (0.4783)&(0.4613)\\
        \textbf{$ListAge$}  &  &         & &  &-0.0010& 0.1365***& 0.5530***&0.1578***\\
                &      &     & &  &(-0.0413)& (3.7156)& (6.5333)&(5.5267)\\
        \textbf{$Lev$}     &   &        & &  &0.0536& -0.0620& 6.7446***&0.0894*\\
                &      &     & &  &(0.8913)& (-0.9460)& (32.842)&(1.7679)\\
        \textbf{$Size$}    &     &      & &  &0.6722***& 0.4376***& -0.4512***&0.4410***\\
                &    &       & &  &(49.844)& (27.508)& (-9.5683)&(35.103)\\
        \textbf{$Avg$}     &     &      & &  &-0.1597***& -0.0907***& -0.5288***&-0.0739***\\
                &      &     & &  &(-14.054)& (-7.0579)& (-13.531)&(-7.5015)\\
        \textbf{$Der$}     &     &      & &  &-0.0256***& -0.0128& 0.0191&-0.0184***\\
                &      &     & &  &(-3.1008)& (-1.5259)& (0.6820)&(-2.8025)\\
        \midrule
        \textbf{Controls} & NO &NO& NO&  NO&YES   & YES& YES  &YES\\
        \textbf{Year} & YES  &YES   & YES       &  YES&YES  & YES& YES  &YES\\
        \textbf{Id} & YES  &YES   & YES       &  YES&YES  & YES& YES  &YES\\
        \textbf{$N$}   & 10461&10461& 12595&  12595&10461& 10461& 12595&12595\\
        \textbf{$R^2$}    & 0.0011&0.0011& 0.0010&  0.0011&0.2790& 0.1145& 0.4112&0.1511\\
        \bottomrule
        \end{tabular}
        \begin{tablenotes}
        \item Note: Similar to the robustness tests, $Dis$ corresponds to regressions on ESG rating disagreement in year t and human capital in year t+1, while $Disb$ corresponds to regressions on ESG rating disagreement in year t-1 and human capital in year t.
        \end{tablenotes}
        \end{threeparttable}
    \end{table}
    \end{landscape}
    
    \begin{landscape}
        \begin{table}[H]
            \centering
            \caption{Heterogeneity Analysis}
            \small
            \label{tab:Hethero}
            \begin{threeparttable}

            \begin{tabular}{lp{1.8cm} p{1.8cm} p{1.8cm} p{2cm} p{1.8cm} p{2cm}cc}
                \toprule
            \cmidrule(lr){2-9}
            \textbf{Variables}      & State-Owned & Non-State-Owned & Capital-Intensive & Non-Capital-Intensive & High-Pollution & Low-Pollution  & High-Tech&Non-High-Tech\\
                \midrule
                \textbf{} & \multicolumn{8}{c}{\textbf{$TFP_{LP}$}}  \\
                \midrule
                $Dis$ & -0.0622** & -0.0310* & -0.0322 & -0.0462*** & -0.0239 & -0.0481***  & -0.0425**&-0.0043\\
                   & (-2.0516) & (-1.8294) & (-1.1009) & (-2.8884) & (-1.0401) & (-2.7959)  & (-2.2163)&(-0.1653)\\
                \multirow{1}{*}{\textbf{Chow Test}} & \multicolumn{2}{c}{500.19***} & \multicolumn{2}{c}{289.12***} & \multicolumn{2}{c}{544.57***} &  \multicolumn{2}{c}{756.47***}\\
                \textbf{$R^2$} & 0.1573 & 0.1886 & 0.1741 & 0.1817 & 0.2010 & 0.1754  & 0.1751&0.1570\\
                \midrule

                     \textbf{} & \multicolumn{8}{c}{\textbf{$TFP_{OP}$}}  \\
                \midrule
                $Dis$ & -0.0553* & -0.0268 & -0.0097 & -0.0458*** & -0.0070 & -0.0493**  & -0.0365**&-0.0048\\
                   & (-1.8608) & (-1.5208) & (-0.3229) & (-2.6338) & (-0.3149) & (-2.4735)  & (-2.2763)&(-0.1639)\\
                \multirow{1}{*}{\textbf{Chow Test}} & \multicolumn{2}{c}{529.58***} & \multicolumn{2}{c}{269.31***} & \multicolumn{2}{c}{534.66***} &  \multicolumn{2}{c}{760.50***}\\
                \textbf{$R^2$} & 0.1759 & 0.1593 & 0.1672 & 0.1663 & 0.1930 & 0.1595  & 0.1561&0.1526\\
                \midrule

                        \textbf{} & \multicolumn{8}{c}{\textbf{$TFP_{GMM}$}}  \\
                \midrule
                $Dis$ & -0.0618** & -0.0310* & -0.0278 & -0.0471*** & -0.0134 & -0.0519**  & -0.0413**&-0.0122\\
                   & (-2.1381) & (-1.6745) & (-0.9906) & (-2.6889) & (-0.6185) & (-2.5484)  & (-2.4165)&(-0.4205)\\
                \multirow{1}{*}{\textbf{Chow Test}} & \multicolumn{2}{c}{498.97***} & \multicolumn{2}{c}{248.15***} & \multicolumn{2}{c}{475.42***} &  \multicolumn{2}{c}{698.54***}\\
                \textbf{$R^2$} & 0.1607 & 0.1254 & 0.1504 & 0.1377 & 0.1691 & 0.1321  & 0.1256&0.1315\\
                \midrule
                \textbf{Controls} & YES & YES & YES & YES & YES & YES  & YES &YES  
        \\
                \textbf{Year} & YES & YES & YES & YES & YES & YES  & YES &YES  
        \\
                \textbf{Id} & YES & YES & YES & YES & YES & YES  & YES &YES  
        \\
                \textbf{Double Clustering} & YES & YES & YES & YES & YES & YES  & YES &YES  \\
        
                \textbf{$N$} & 4416 & 9001 & 2342 & 11075 & 3900 & 9517  & 8454&4963\\
        
                \bottomrule
            \end{tabular}
            \begin{tablenotes}
                \item Note: This table presents the results of a heterogeneity analysis on the negative impact of ESG rating divergence on TFP, confirming that the negative effect is more pronounced in state-owned, non-capital-intensive, low-pollution, and high-tech firms.
            \end{tablenotes}       
        \end{threeparttable}
        \end{table}
        \end{landscape}
        \begin{table}[H]
                \centering
                \caption{Performance Metrics for Training and Test Sets}
                \label{tab:calsusal}
                \begin{threeparttable}
                    \begin{tabular}{c p{1.5cm} p{1.5cm} p{1.5cm} p{1.5cm} |p{1.5cm} p{1.5cm} p{1.5cm} p{1.5cm}}
                    \toprule
                    \textbf{Metric} & \textbf{Training Set 1} & \textbf{Training Set 2} & \textbf{Training Set 3} & \textbf{Training Set 4} & \textbf{Test Set 1} & \textbf{Test Set 2} & \textbf{Test Set 3} & \textbf{Test Set 4} \\ 
                    \midrule
                    \textbf{MSE}  & 0.2609           & \textbf{0.2274}           & 0.2818             & 0.4714           & 0.3492           & \textbf{0.3300}           & 0.3835           & 0.4728           \\
                    \textbf{RMSE} & 0.5108           & \textbf{0.4769}           & 0.5309             & 0.6866           & 0.5909           & \textbf{0.5744}           & 0.6192           & 0.6876           \\
                    \textbf{MAE}  & 0.3995           & \textbf{0.3729}           & 0.4145             & 0.5383           & 0.4589           & \textbf{0.4523}           & 0.4818           & 0.5415           \\
                    \textbf{MAPE} & 4.7786\%          & \textbf{4.4569\%}          & 4.9457\%            & 6.4047\%          & 5.4649\%          & \textbf{5.3644\%}          & 5.7293\%          & 6.4399\%          \\
                    \textbf{R\textsuperscript{2}}    & 0.7685           & \textbf{0.7969}           & 0.7500             & 0.5794           & 0.7045           & \textbf{0.7176}           & 0.6755           & 0.5955           \\
                    \bottomrule
                    \end{tabular}
                    \begin{tablenotes}
                        \footnotesize
                        \item Note: Set1 is the XGBoost performance of $Dis$ in the same period as the TFP, Set2 is the XGBoost performance of $Dis$ in the preceding period, Set3 is the XGBoost performance excluding $Dis$, and Set4 is the model performance of the linear model (no fixed effects) and $Dis$ in the preceding period.
                    \end{tablenotes}
                \end{threeparttable}
            \end{table}
            
  \clearpage
            
            \clearpage
            
 \appendix        
\section{Online Appendix}

\subsection{Descriptive Statistics }
\begin{table}[H]
    \centering
    \caption{Descriptive Statistics of Main Variables}
    \begin{threeparttable}
    \begin{tabular}{lcccccccc}
    \toprule
    \textbf{Variables} & \textbf{Count} & \textbf{Mean} & \textbf{SD} & \textbf{Min} & \textbf{25\%} & \textbf{50\%} & \textbf{75\%} & \textbf{Max} \\
    \midrule
    $TFP_{LP}$ & 13417 & 8.51 & 1.04 & 6.38 & 7.79 & 8.40 & 9.14 & 11.28 \\
    $TFP_{OP}$ & 13417 & 6.83 & 0.85 & 5.14 & 6.24 & 6.72 & 7.33 & 9.17 \\
    $TFP_{GMM}$ & 13417 & 5.72 & 0.80 & 4.09 & 5.18 & 5.63 & 6.18 & 8.12 \\
    $Dis$ & 13417 & 0.26 & 0.19 & 0.01 & 0.11 & 0.24 & 0.34 & 0.83 \\
    $TobinQ$ & 13417 & 2.03 & 1.32 & 0.83 & 1.22 & 1.61 & 2.31 & 8.34 \\
    $Top10$ & 13417 & 56.77 & 15.15 & 23.59 & 45.91 & 56.95 & 67.87 & 90.89 \\
    $ListAge$ & 13417 & 2.34 & 0.68 & 1.10 & 1.79 & 2.40 & 3.00 & 3.37 \\
    $Lev$ & 13417 & 0.43 & 0.19 & 0.06 & 0.28 & 0.43 & 0.57 & 0.88 \\
    $Size$ & 13417 & 22.52 & 1.34 & 20.09 & 21.54 & 22.32 & 23.31 & 26.41 \\
    $Avg$ & 13417 & 4.79 & 0.80 & 3.07 & 4.25 & 4.69 & 5.25 & 7.17 \\
    $Der$ & 13417 & 1.05 & 1.07 & 0.07 & 0.39 & 0.74 & 1.31 & 6.75 \\
    $EI$ & 13417 & 1.30 & 1.49 & 0.00 & 0.00 & 1.10 & 2.30 & 7.74 \\
    $WW$ & 10059 & -1.03 & 0.08 & -1.24 & -1.08 & -1.03 & -0.98 & -0.86 \\
    $KZ$ & 13417 & 1.17 & 1.94 & -4.35 & 0.02 & 1.36 & 2.49 & 6.22 \\
    $Master$ & 10461 & 4.31 & 1.55 & 1.10 & 3.18 & 4.20 & 5.31 & 8.42 \\
    $Undergrad$ & 12595 & 6.31 & 1.35 & 3.56 & 5.35 & 6.17 & 7.15 & 10.05 \\
    \bottomrule
    \end{tabular}
    \begin{tablenotes}
    \item Note: The data in the table are descriptive statistics for deleting missing values after lagging one period of TFP.
    \end{tablenotes}
    \end{threeparttable}
\end{table}
\subsection{Correlation Coefficient}
\begin{landscape}

\begin{table}[b]
    \centering
    \caption{Correlation Matrix}
    \small
    \begin{threeparttable}

    \begin{tabular}{lrrrrrrrrcrrrr}
    \toprule
    \textbf{} & \textbf{$TFP_{LP}$} & \textbf{$TFP_{OP}$} & \textbf{$TFP_{GMM}$} & \textbf{$Dis$} & \textbf{$TobinQ$} & \textbf{$Top10$} & \textbf{$ListAge$} & \textbf{$Lev$} & \textbf{$Size$} & \textbf{$Avg$} & \textbf{$Der$} & \textbf{$N.TFP_{province}$} & \textbf{$IV$} \\
    \midrule
    \textbf{$TFP_{LP}$} & 1.000 & 0.951 & 0.918 & -0.066 & -0.200 & 0.127 & 0.329 & 0.444 & 0.801 & 0.648 & 0.344 & -0.037 & 0.082 \\
    \textbf{$TFP_{OP}$} & 0.951 & 1.000 & 0.976 & -0.054 & -0.183 & 0.104 & 0.303 & 0.405 & 0.707 & 0.841 & 0.325 & -0.036 & 0.067 \\
    \textbf{$TFP_{GMM}$} & 0.918 & 0.976 & 1.000 & -0.041 & -0.132 & 0.080 & 0.244 & 0.356 & 0.583 & 0.823 & 0.295 & -0.005 & 0.059 \\
    \textbf{$Dis$} & -0.066 & -0.054 & -0.041 & 1.000 & 0.014 & -0.078 & 0.025 & 0.039 & -0.071 & -0.024 & 0.052 & -0.015 & -0.974 \\
    \textbf{$TobinQ$} & -0.200 & -0.183 & -0.132 & 0.014 & 1.000 & -0.036 & -0.094 & -0.249 & -0.283 & -0.116 & -0.173 & 0.012 & -0.008 \\
    \textbf{$Top10$} & 0.127 & 0.104 & 0.080 & -0.078 & -0.036 & 1.000 & -0.364 & -0.066 & 0.147 & 0.044 & -0.052 & 0.038 & 0.087 \\
    \textbf{$ListAge$} & 0.329 & 0.303 & 0.244 & 0.025 & -0.094 & -0.364 & 1.000 & 0.292 & 0.424 & 0.204 & 0.240 & -0.184 & -0.024 \\
    \textbf{$Lev$} & 0.444 & 0.405 & 0.356 & 0.039 & -0.249 & -0.066 & 0.292 & 1.000 & 0.458 & 0.256 & 0.843 & -0.044 & -0.038 \\
    \textbf{$Size$} & 0.801 & 0.707 & 0.583 & -0.071 & -0.283 & 0.147 & 0.424 & 0.458 & 1.000 & 0.413 & 0.359 & -0.104 & 0.082 \\
    \textbf{$Avg$} & 0.648 & 0.841 & 0.823 & -0.024 & -0.116 & 0.044 & 0.204 & 0.256 & 0.413 & 1.000 & 0.222 & -0.041 & 0.027 \\
    \textbf{$Der$} & 0.344 & 0.325 & 0.295 & 0.052 & -0.173 & -0.052 & 0.240 & 0.843 & 0.359 & 0.222 & 1.000 & -0.047 & -0.046 \\
    \textbf{$N.TFP_{province}$} & -0.037 & -0.036 & -0.005 & -0.015 & 0.012 & 0.038 & -0.184 & -0.044 & -0.104 & -0.041 & -0.047 & 1.000 & 0.014 \\
    \textbf{$IV$} & 0.082 & 0.067 & 0.059 & -0.974 & -0.008 & 0.087 & -0.024 & -0.038 & 0.082 & 0.027 & -0.046 & 0.014 & 1.000 \\
    \bottomrule
    \end{tabular}
\begin{tablenotes}
        \item Note: This is the matrix of linear correlation coefficients for all variables.
\end{tablenotes}        
\end{threeparttable}
\end{table}
\end{landscape}
\subsection{Robustness Tests and Endogeneity Treatments}
To ensure the robustness of the baseline regression results, the following robustness tests were conducted.

First, to verify whether Hypothesis H1 holds when only leading $Dis$ by one period, we lead the it by one period(see \autoref{tab:robustness_test1}).

Second, since the TFP of firms in this study is influenced by regional productivity levels, to mitigate endogeneity issues arising from omitted variable bias, we include the provincial-level New quality productivity measured by \citet{LuJiang} as a control variable (see \autoref{tab:robustness_test2}).

Third, to further mitigate everse causality endogeneity issues, following \citet{YangJinyu}, we calculate the industry average of ESG rating disagreements by year and industry, and use the cube of the difference between the industry average and the ESG rating disagreements as an instrumental variable. A two-stage least squares (2SLS) estimation is conducted, and the Kleibergen-Paap rk LM statistics verify the verify the validity of instrumental variables(see \autoref{tab:robustness_test3}).

Fourth, to address sample selection bias, we denote disagreements under 20\% as non-existent and the remainder as present. Propensity scores are computed via logistic regression, followed by 1:1 nearest-neighbor matching and subsequent regression (see \autoref{tab:robustness_test4}).
\begin{table}[H]
    \centering
    \caption{Robustness Tests (I)}
    \label{tab:robustness_test1}
    \begin{threeparttable}
    \begin{tabular}{lcccccc}
    \toprule
    \textbf{Variables} & \multicolumn{6}{c}{\textbf{TFP}}  \\ 
    \cmidrule(lr){2-7}
    &\textbf{$TFP_{LP}$} &\textbf{$TFP_{OP}$} &\textbf{$TFP_{GMM}$} & \textbf{$TFP_{LP}$}& \textbf{$TFP_{OP}$}& \textbf{$TFP_{GMM}$} \\ 
    \midrule
    \textbf{$Dis$}     & -0.0662*** &-0.0524** &-0.0524** & -0.0246**  & -0.0159** & -0.0180** \\ 
            & (-2.7472)  & (-2.2660)&(-2.2781) & (-2.2085)  & (-2.4677) & (-2.0769) \\
    \textbf{$TobinQ$}  &           & & & 0.0218***   & 0.0141*** & 0.0150*** \\
            &           & & & (6.8126)    & (7.0455)  & (6.1013) \\
    \textbf{$Top10$}   &           & & & 0.0005      & 0.0002    & 0.0004 \\
            &           & & & (0.8363)    & (0.5950)  & (0.8865) \\
    \textbf{$ListAge$}  &           & & & -0.0510**   & -0.0291** & -0.0678*** \\
            &           & & & (-2.3362)   & (-2.2563) & (-4.0705) \\
    \textbf{$Lev$}     &           & & & 0.1415      & 0.0739    & 0.0650 \\ 
            &           & & & (1.8712)    & (1.5504)  & (1.2862) \\
    \textbf{$Size$}    &           & & & 0.4212***   & 0.2448*** & 0.1800*** \\
            &           & & & (21.576)    & (18.539)  & (11.207) \\
    \textbf{$Avg$}     &           & & & 0.7135***   & 0.8172*** & 0.8193*** \\
            &           & & & (38.466)    & (64.699)  & (55.668) \\
    \textbf{$Der$}     &           & & & -0.0257**   & -0.0113   & -0.0135 \\
            &           & & & (-2.0169)   & (-1.3877) & (-1.5955) \\
    \midrule
    \textbf{Controls} & NO &NO &NO & YES & YES & YES \\
    \textbf{Year} & YES     & YES&YES & YES       & YES & YES \\
    \textbf{Id} & YES     &YES &YES & YES       & YES & YES \\
    \textbf{Doubel Clustering}& YES     &YES &YES & YES  & YES & YES \\
    \textbf{$N$}   & 13417   &13417 &13417 & 13417     & 13417 & 13417 \\
    \textbf{$R^2$}    & 0.0015  &0.0011 &0.0011 & 0.7027    & 0.8496 & 0.7734 \\
    \bottomrule
    \end{tabular}
    \begin{tablenotes}
        \item Note: This is the result of only front-loading $Dis$ compared to $TFP$ by one period. In this regression result, $Dis$ denotes ESG Rating Disagreement of $t-1$ year and $TFP$ denotes total factor productivity of $t$ year. Unless otherwise noted, $Dis$ and $TFP$ are consistent with the baseline regression settings.
        \end{tablenotes}
    \end{threeparttable}
\end{table}
\begin{table}[H]
    \centering
    \caption{Robustness Tests (II)}
    \label{tab:robustness_test2}
    \begin{threeparttable}
    \begin{tabular}{lcccccc}
    \toprule
    \textbf{Variables} & \multicolumn{6}{c}{\textbf{TFP}} \\ 
    \cmidrule(lr){2-7}
    &\textbf{$TFP_{LP}$} &\textbf{$TFP_{OP}$} & \textbf{$TFP_{GMM}$}& \textbf{$TFP_{LP}$}& \textbf{$TFP_{OP}$}& \textbf{$TFP_{GMM}$} \\ 
    \midrule
    \textbf{$Dis$}     & -0.0647** & -0.0519**&-0.0510** & -0.0419***  & -0.0342** & -0.0346** \\ 
            & (-2.5465)  &(-1.9984) &(-2.0263) & (-2.6444)  & (-1.9838) & (-1.9658) \\
    \textbf{$TobinQ$}  &           & & & -0.0027   & -0.0013 & -0.0003 \\
            &           & & & (-0.4030)  & (-0.2619) & (-0.0725) \\
    \textbf{$Top10$}   &           & & & -0.0009   & -0.0009 & -0.0005 \\
            &           & & & (-0.9777)  & (-0.9697) & (-0.5039) \\
    \textbf{$ListAge$}  &           & & & 0.0906**   & 0.0796** & 0.0058 \\
            &           & & & (2.4895)   & (2.3720) & (0.1639) \\
    \textbf{$Lev$}     &           & & & -0.1555    & -0.1408 & -0.1334 \\ 
            &           & & & (-1.3079)  & (-1.2639) & (-1.1864) \\
    \textbf{$Size$}    &           & & & 0.3762***   & 0.2596*** & 0.2262*** \\
            &           & & & (14.285)    & (13.442) & (10.036) \\
    \textbf{$Avg$}     &           & & & 0.2277***   & 0.2680** & 0.2660** \\
            &           & & & (2.6951)    & (2.7597) & (2.6807) \\
    \textbf{$Der$}     &           & & & -0.0210*   & -0.0194 & -0.0217 \\
            &           & & & (-1.9537)   & (-1.7844) & (-1.9273) \\
    \textbf{$N.TFP_{province}$}  &           & & & 0.0993     & 0.0847 & 0.0403 \\
            &           & & & (0.6302)    & (0.5857) & (0.2531) \\
    \midrule
    \textbf{Controls} & NO &NO & NO& YES & YES & YES \\
    \textbf{Year} & YES     & YES& YES& YES       & YES & YES \\
    \textbf{Id} & YES     & YES&YES & YES       & YES & YES \\
    \textbf{Doubel Clustering}& YES  &YES & YES& YES  & YES & YES \\
    \textbf{$N$}   & 12893   & 12893&12893 & 12893     & 12893 & 12893 \\
    \textbf{$R^2$}    & 0.0014  &0.0011 &0.0010 & 0.1793    & 0.1670 & 0.1406 \\
    \bottomrule
    \end{tabular}
    \begin{tablenotes}
        \item Note: This is the result of the regression with the addition of control variables, and the other settings are consistent with the baseline regression. $N.TFP_{province}$ represents the level of total factor productivity at the provincial level.
        \end{tablenotes}
    \end{threeparttable}
\end{table}
\begin{table}[H]
    \centering
    \caption{Robustness Tests (III)}
    \label{tab:robustness_test3}
    \begin{threeparttable}
    \begin{tabular}{lcccc}
    \toprule
    \textbf{Variables} & \multicolumn{4}{c}{\textbf{Dependent Variable}} \\
    \cmidrule(lr){2-5}
    & \textbf{$Dis$} & \textbf{$TFP_{LP}$} & \textbf{$TFP_{OP}$} & \textbf{$TFP_{GMM}$} \\
    \midrule
    \textbf{$IV(\hat{Dis})$} & -0.3302*** & -0.0476*** & -0.0409** & -0.0425** \\
    & (-328.74) & (-3.0292) & (-2.3843) & (-2.3456) \\
    \textbf{$TobinQ$} & 0.0002 & -0.0027 & -0.0013 & -0.0003 \\
    & (0.1447) & (-0.3986) & (-0.2574) & (-0.0685) \\
    \textbf{$Top10$} & 0.0001 & -0.0009 & -0.0009 & -0.0005 \\
    & (1.1390) & (-0.9769) & (-0.9679) & (-0.5016) \\
    \textbf{$ListAge$} & -0.0118 & 0.0906** & 0.0797** & 0.0058 \\
    & (-1.2204) & (2.4886) & (2.3696) & (0.1655) \\
    \textbf{$Lev$} & -0.0061 & -0.1555 & -0.1408 & -0.1335 \\
    & (-0.7749) & (-1.3069) & (-1.2627) & (-1.1853) \\
    \textbf{$Size$} & -0.0044** & 0.3761*** & 0.2594*** & 0.2260*** \\
    & (-2.0583) & (14.306) & (13.451) & (10.030) \\
    \textbf{$Avg$} & -0.0015 & 0.2277*** & 0.2680** & 0.2660** \\
    & (-0.5293) & (2.6949) & (2.7600) & (2.6812) \\
    \textbf{$Der$} & 0.0045** & -0.0210* & -0.0193 & -0.0216 \\
    & (3.1346) & (-1.9398) & (-1.7686) & (-1.9099) \\
    \textbf{$N.TFP_{province}$} & -0.0030 & 0.0994 & 0.0847 & 0.0404 \\
    & (-0.2042) & (0.6320) & (0.5877) & (0.2540) \\
    \midrule
    \textbf{Controls} & YES & YES & YES & YES \\
    \textbf{Year} & YES & YES & YES & YES \\
    \textbf{ID} & YES & YES & YES & YES \\
    \textbf{Double Clustering} & YES & YES & YES & YES \\
    \textbf{$F$} & 2.049e+04 & 227.20 & 208.52 & 170.30 \\
    \textbf{$N$} & 12892 & 12892 & 12892 & 12892 \\
 \textbf{Kleibergen-Paap rk LM}& \multicolumn{4}{c}{622.33****}\\
    \textbf{R\(^2\)} & 0.9517 & 0.1794 & 0.1671 & 0.1408 \\
    \bottomrule
    \end{tabular}
    \begin{tablenotes}
    \item \textbf{Note:} This is the result of an instrumental variable test where $TFP$ is the t+1 period, just like the baseline regression.
    \end{tablenotes}
    \end{threeparttable}
\end{table}
\begin{table}[H]
    \centering
    \caption{Robustness Tests (IV)}
    \label{tab:robustness_test4}
    \begin{threeparttable}
    \begin{tabular}{lcccccc}
    \toprule
    \textbf{Variables} & \multicolumn{4}{c}{\textbf{TFP}} & & \\ 
    \cmidrule(lr){2-7}
    & \textbf{$TFP_{LP}$} &\textbf{$TFP_{OP}$} &\textbf{$TFP_{GMM}$} & \textbf{$TFP_{LP}$} & \textbf{$TFP_{OP}$} & \textbf{$TFP_{GMM}$} \\ 
    \midrule
    \textbf{$Dis$} & -0.1054*** &-0.0959*** &-0.0908*** & -0.0730** & -0.0676** & -0.0645** \\ 
    & (-3.0359) &(-2.9849) & (-2.7841)& (-2.3040) & (-2.2900) & (-2.1303) \\
    \textbf{$TobinQ$} & & & & 0.0007 & 0.0002 & -0.0026 \\
    & & & & (0.0934) & (0.0351) & (-0.3889) \\
    \textbf{$Top10$} & & & & 0.0005 & 0.0004 & 0.0007 \\
    & & & & (0.5206) & (0.3886) & (0.6853) \\
    \textbf{$ListAge$} & & & & 0.1119 & 0.1001 & 0.0149 \\
    & & & & (1.6127) & (1.5470) & (0.2245) \\
    \textbf{$Lev$} & & & & -0.0295 & -0.0270 & -0.0174 \\ 
    & & & & (-0.3276) & (-0.3216) & (-0.2021) \\
    \textbf{$Size$} & & & & 0.3899*** & 0.2685*** & 0.2291*** \\
    & & & & (16.968) & (12.538) & (10.440) \\
    \textbf{$Avg$} & & & & 0.2343*** & 0.2761*** & 0.2832*** \\
    & & & & (13.111) & (16.582) & (16.598) \\
    \textbf{$Der$} & & & & -0.0408*** & -0.0368*** & -0.0385*** \\
    & & & & (-3.4013) & (-3.2928) & (-3.3577) \\
    \midrule
    \textbf{Controls} & NO &NO &NO & YES & YES & YES \\
    \textbf{Year} & YES &YES &YES & YES & YES & YES \\
    \textbf{ID} & YES & YES&YES & YES & YES & YES \\
    \textbf{Double Clustering} & NO &NO &NO & NO & NO & NO \\
    \textbf{$N$} & 5368 &5368 &5368 & 5368 & 5368 & 5368 \\
    \textbf{R\(^2\)} & 0.0028 &0.0027 &0.0024 & 0.1739 & 0.1628 & 0.1455 \\ 
    \bottomrule
    \end{tabular}
    \begin{tablenotes}
        \item Note: This is the result after PSM, matched on a 1:1 basis, with other settings not differing from the baseline regression.
    \end{tablenotes}
    \end{threeparttable}
\end{table}
\subsection{Optuna Parameter Range}
\begin{table}[H]
\centering
\caption{Hyperparameters for XGBoost Model}
\begin{threeparttable}

\begin{tabular}{ll}
\toprule
\textbf{Parameter}      & \textbf{Range(optuna) / Value}       \\ 
\midrule
\texttt{loss\_function}      & \texttt{rmse}    \\ 
\texttt{learning\_rate} & [0.005, 0.2] (log scale)     \\ 
\texttt{max\_depth}     & [1, 3]                    \\ 
\texttt{min\_child\_weight} & [1, 5] (log scale)       \\ 
\texttt{subsample}      & [0.5, 0.8]                  \\ 
\texttt{colsample\_bytree} & [0.5, 0.8]               \\ 
\texttt{reg\_lambda}    & [1, 10] (log scale)         \\ 
\texttt{reg\_alpha}     & [1, 10] (log scale)         \\ 
\texttt{tree\_method}   & \texttt{hist}               \\ 
\texttt{device}         & \texttt{cuda}               \\ 
\texttt{num\_boost\_round} & 1000                    \\ 
\texttt{nfold}          & 20                          \\ 
\texttt{early\_stopping\_rounds} & 5                 \\ 
\texttt{seed}           & 42                          \\ 
\bottomrule
\end{tabular}
\begin{tablenotes}
        \item Note: This table shows the range of hyperparameters Optuna searches for, automatically selecting the optimal hyperparameters for the final model configuration.
\end{tablenotes}
\end{threeparttable}
\end{table}

\end{document}